\documentclass[preprintnumbers,superscriptaddress,nofootinbib,aps,prd,floatfix]{revtex4}

\usepackage{hyperref}
\usepackage{amsmath,slashed}
\usepackage{graphicx}
\usepackage{subfigure}

\newcommand{\be}{\begin{eqnarray*}}
\newcommand{\ee}{\end{eqnarray*}}

\newcommand{\bee}{\begin{eqnarray}}
\newcommand{\eee}{\end{eqnarray}}
\newcommand{\beeq}{\begin{equation}}
\newcommand{\eeeq}{\end{equation}}

\newcommand{\mc}{\mathcal}

\newcommand{\maps}{\textsc{hytrees}}
\newcommand{\dire}{\textsc{Dire}\;}
\newcommand{\pythia}{\textsc{Pythia}\;}

\newcommand{\me}[2]{\left|\mathcal{M}\left({#1}\right)^{#2}\right|^2}

\DeclareGraphicsExtensions{.pdf,.png,.jpg}

\usepackage{color}

\begin{document}

\title{HYTREES: Combining Matrix Elements and Parton Shower for Hypothesis Testing}

\begin{abstract}
We present a new way of performing hypothesis tests on scattering data, by means of a perturbatively calculable classifier. This classifier exploits the ``history tree" of how the measured data point might have evolved out of any simpler (reconstructed) points along classical paths, while explicitly keeping quantum-mechanical interference effects by copiously employing complete leading-order matrix elements. This approach extends the standard Matrix Element Method to an arbitrary number of final state objects and to exclusive final states where reconstructed objects can be collinear or soft. We have implemented this method into the standalone package \maps~and have applied it to Higgs boson production in association with two jets, with subsequent decay into photons. \maps~allows to construct an optimal classifier to discriminate this process from large Standard Model backgrounds. It further allows to find the most sensitive kinematic regions that contribute to the classification.
\end{abstract}

\author{Stefan Prestel}
\affiliation{Department of Astronomy and Theoretical Physics,\\Lund University, S-223 62 Lund, Sweden}

\author{Michael Spannowsky}
\affiliation{Institute for Particle Physics Phenomenology, Department
  of Physics,\\Durham University, DH1 3LE, United Kingdom}

\preprint{IPPP/19/1}
\preprint{LU-TP 19-07}
\vspace*{4ex}

\maketitle

\section{Introduction}
\label{sec:intro}

The separation of interesting signal events from large Standard-Model induced
backgrounds is one of the biggest challenges in searches for new physics and 
when measuring particle properties at the LHC.
This problem is magnified when the final-states of interest 
have a large probability to be produced proton-proton collisions 
according to the Standard Model. Typical classifications into signal and 
background events are based on observables that are characteristic of the the 
quantum numbers of the particles involved in each hypothesis. For example, the quantum
numbers (e.g. charges, spin and mass) of a resonance result in a specific 
radiation profile in the detector. The radiation induced by such a 
resonance is more likely to populate specific phase space regions. Thus, to
infer if a process is induced by signal or by background, one wants to know 
how likely the measured radiation profile was induced by either 
hypothesis, i.e. $\mathcal{P}(\{p_i\}|S)$ for signal 
and $\mathcal{P}(\{p_i\}|B)$ for background, where $\{p_i\}$ denotes the set 
of 4-momenta measured in the detector. The Neyman-Pearson Lemma shows 
\cite{James:2000et} that by taking the ratio between both probabilities
\begin{equation}
\label{eq:chi}
\chi = \frac{\mathcal{P}(\{p_i\}|S)}{\mathcal{P}(\{p_i\}|B)}
\end{equation}
yields an ideal classifier. This approach underlies the so-called Matrix 
Element Method (MEM) \cite{Kondo:1988yd}, which has been used in a large 
variety of contexts \cite{Abazov:2004cs, Abulencia:2005pe, Cranmer:2006zs,
Gao:2010qx, Andersen:2012kn, Martini:2015fsa, Gritsan:2016hjl}. In the
MEM, the probabilities $\mathcal{P}(\{p_i\}|S)$ and $\mathcal{P}(\{p_i\}|B)$ 
are calculated directly from the matrix elements of the respective ``hard" 
processes. In \cite{Soper:2011cr, Soper:2012pb} the parton-level MEM has 
been extended to including the parton shower in the evaluation of the 
probabilities, and has been implemented in Shower \cite{Soper:2011cr, 
Soper:2012pb, FerreiradeLima:2016gcz} and Event \cite{Soper:2014rya, 
Englert:2015dlp, FerreiradeLima:2017iwx} Deconstruction, thereby allowing
for the analysis of an arbitrary number of final state objects. Information
from the parton shower is particularly important in jet-rich final states and
in the comparison of the substructure of jets for classification. Here exclusive
fixed-order matrix elements do not provide a good description of nature, due to
the appearance of collinear and soft divergences in the matrix elements. 

Conversely, LHC signals and backgrounds are often predicted by using
General-Purpose Event Generators (see e.g.~\cite{Buckley:2011ms}) to produce
scattering event pseudo-data. In this context, several frameworks to combine 
the parton shower with multiple hard matrix elements of the multi-jet processes
containing  have been laid out, e.g.~the MLM \cite{Mangano:2001xp} procedure, 
the CKKW \cite{Catani:2001cc} and CKKW-L \cite{Lonnblad:2001iq} methods, or 
iterated matrix-element corrections approach~\cite{Giele:2011cb,Fischer:2017yja}. 
These formalisms allow to avoid double counting between jets generated during 
the parton shower step and the matrix element, such that multiple jets can 
simultaneously be described with matrix-element accuracy. 

We are proposing to combine techniques used traditionally for the CKKW-L and 
the iterated matrix-element correction approach of~\cite{Fischer:2017yja}, and 
then use the resulting procedure to construct sophisticated perturbative 
weights of the full input event, that will facilitate the classification 
between signal and background. To calculate $\mathcal{P}(\{p_i\}|S)$ and 
$\mathcal{P}(\{p_i\}|B)$, one needs to evaluate all possible combinations of 
parton shower and hard process histories that can give rise to the final 
state $\{p_i\}$.
Conceptually, such an analysis method is suitable for any final state of 
interest consisting of reconstructed objects, i.e. arbitrary numbers of 
isolated leptons, photons and jets. The approach, dubbed \maps, is in line 
with the Shower/Event deconstruction method, but goes beyond these by including 
hard matrix elements with multiple jet emissions to calculate the weights of 
the event histories. We describe here the first implementation of such a method 
and showcase it in the context of a concrete example which is highly relevant 
for Higgs phenomenology, i.e. $pp \to (\mathrm{H}\to \gamma \gamma) + \mathrm{jets}$. 

The outline of the paper is as follows.
In Sec.~\ref{sec:implementation} we discuss the details of the \maps\ algorithm. \maps\ relies
on the \dire parton shower~\cite{Hoche:2015sya} to calculate the weights of the 
event histories. For details on the splitting probabilities used in the \dire 
dipole shower we refer to Appendix~\ref{app:dire}. In Sec.~\ref{sec:results} we 
apply \maps\ to the study of the classification of the process 
$pp \to (\mathrm{H}\to \gamma \gamma) + \mathrm{jets}$ versus the processes without Higgs
boson that lead to $pp \to \gamma \gamma +\mathrm{jets}$. We offer conclusions
in Sec.~\ref{sec:conclusion}. 

\section{Implementation of \maps}
\label{sec:implementation}

\begin{figure*}[hpt!]
\includegraphics[width=1.0\textwidth]{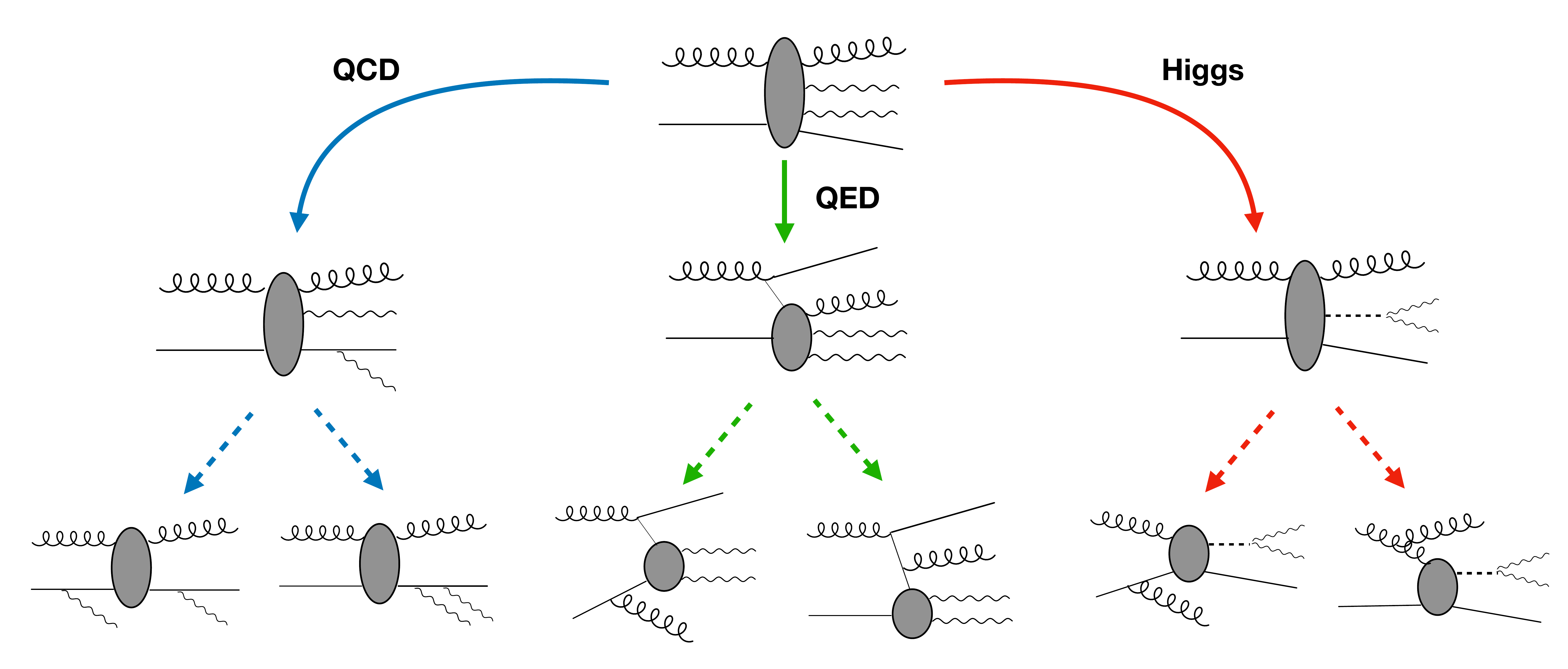}\hspace{0.5cm}
\caption{\label{fig:histories} Pictorial representation of the paths 
contributing to the calculation of the probabilities 
$\mathcal{P}(\{p_i\}|\,\textnormal{Higgs})$, $\mathcal{P}(\{p_i\}|\,\textnormal{QED})$ 
and $\mathcal{P}(\{p_i\}|\,\textnormal{QCD})$, as described in the text.}
\end{figure*}

The definition of the classifier $\chi$ suggested in Eq.~\ref{eq:chi} is 
in principle very intuitive. A practical implementation however requires
assumptions and abstractions before the classifier can be calculated on 
experimental data. Thus, to test and develop the classifier, we will use 
event generator pseudo-data. We will evaluate the new classifier on this
pseudo-data. To be concrete, we use realistic (showered and hadronised) events, i.e. 
each ``event" consists of a collection of particles -- photons, leptons, 
long-lived hadrons, {\it etc.} -- with each particle represented by a 
4-vector stored in \textsc{HepMc} event format \cite{Dobbs:2001ck}. The hard 
process underlying these events were generated using 
MadGraph \cite{Alwall:2014hca}, and showered and hadronised
using \pythia~\cite{Sjostrand:2014zea}.

These events are further processed to arrive at final states consisting of 
reconstructed objects, i.e.~isolated leptons, isolated photons or jets. A
lepton $(e,\mu)$ or photon isolated from hadronic energy by demanding that the
total hadronic activity around in a cone of radius $R=0.3$ around the object must contain less than 
$10\%$ of its $p_T$, and the object is required to have $p_T \geq 20$~GeV and 
$|y|<2.5$. Jets are reconstructed using the anti-kT 
algorithm~\cite{Cacciari:2008gp} as implemented in
\texttt{fastjet}~\cite{Cacciari:2011ma}, with radius $R=0.4$. We only consider 
events with at least two jets of $p_{T,j} \geq 35$~GeV, since looser cuts are 
usually not considered in experimental analyses at the LHC. 
After these steps, the final state of interest is now considerably simplified 
compared to the particle-level final state, only consisting of 
$\mathcal{O}(10)$ reconstructed objects. On these states, we will want to 
calculate $\chi$ of Eq.~\ref{eq:chi} from first principles relying
on perturbative methods. Thus, we want to be as insensitive as possible from 
experimental or non-perturbative effects, such as hadronisation or 
pileup-induced soft scatterings. Using reconstructed objects as input to our 
calculation protects us to a large degree from contributions that are 
theoretically poorly controllable.

To allow the calculation of the 
classifier to be as detailed and physical as possible, we will directly use a 
parton shower to calculate the necessary factors. For this, we identify the 
reconstructed objects in the event with partons of a parton shower, i.e. with 
the perturbative part of the event generation before hadronisation. The first 
necessary step is to redistribute 
momenta to ensure that all jet momenta can be mapped to on-shell parton momenta,
and then adding beam momenta defined by momentum conservation in the 
center-of-mass frame. Each of these events is then
translated to \emph{all possible} partonic pseudo-events, by assigning all possible 
parton flavors and all possible color connections to the 
jets\footnote{We want to thank Valentin Hirschi for collaboration at an
early stage of this project, and in particular for sharing a private code
to generate all color connections in a parton ensemble.}. The resulting collection
of events are then passed to the parton shower algorithm\footnote{These events are stored in Les Houches
event files~\cite{Alwall:2006yp}, and read by \textsc{Pythia}, which acts
also as an interface to the parton shower.} to calculate all
necessary weights.

The general philosophy is illustrated in Fig~\ref{fig:histories}.
A reasonable probability for the six configurations in the lowest layer
should depend on the $2\rightarrow 2$ matrix elements for particles connected
to the ``hard" scattering (grey blob). At the same time, the probability of
the three configurations in the middle layer should be proportional
to the $2\rightarrow 3$ matrix elements for particles connected to
the blob, and the overall probability of the top layer should be proportional
to the $2\rightarrow 4$ matrix elements. It is crucial to keep these conditions
in mind when attempting a classification, since in general, the distinction
between ``hard scattering" and ``subsequent radiation" is only well-defined
in the phase-space region of ordered soft- and or collinear emissions. In such
phase-space regions, the quantum-mechanical amplitudes factorize into 
independent building blocks (such as splitting functions or eikonal factors) 
that effectively make up a ``classical" path. If the kinematics of
the event is such that interference effects between the amplitudes for
different paths (i.e. hypotheses) are sizable, then this needs to be
reflected in the classifier. There should not be any discriminating power
for such events. Here, we will build a classifier that \emph{does depend on assigning
a classical path} to phase-space points. The kinematics of each unique point
will be used to calculate the rate of classical paths, such as the ones 
illustrated in Fig~\ref{fig:histories}. In phase-space regions that
allow a (quantum-mechanically) sensible discrimination, the rates of the
dominant paths will factorize into products of squared low-multiplicity 
matrix elements and approximate (splitting) kernels. In all other regions,
we should be as agnostic as possible to the path. These two regions
can be reconciled by always using the complete, non-factorized matrix 
elements to calculate the rate, and only employ the approximate (splitting) kernels
to ``project out" the rate of paths. This will guarantee that we minimize the 
dependence of assigning classical paths in inappropriate phase-space regions.
We can succeed in defining the rate by the full non-factorized matrix element,
for events of varying multiplicity, by employing the iterated matrix-element 
correction probabilities derived in~\cite{Fischer:2017yja} (see Eq.~(15) therein)
when calculating the probability of each path. The simultaneous use of 
matrix-elements for different multiplicities is a significant improvement over 
traditional matrix-element methods, which only leverage matrix-elements for a 
fixed multiplicity at a time. 

The calculation of the classifier thus proceeds by constructing all possible ways 
how the partonic input state could have evolved out of a sequence of 
lower-multiplicity partonic states, by explicitly constructing all 
lower-multiplicity intermediate states via successive recombination of 
three into two particles, until no further recombination is possible. This
construction of all ``histories" follows closely the ideas used in 
matrix-element and parton shower merging methods~\cite{Lonnblad:2001iq}.
The probability of an individual recombination sequence relies on 
full matrix elements as much as possible. In particular, we ensure
that not only the probability of the lowest-multiplicity state is given
by leading-order matrix elements, but that the probability of 
higher-multiplicity states is simultaneously determined by leading-order matrix
elements. Further improvements of the method to incorporate running coupling 
effects, rescaling of parton distributions due to changes in initial-state 
longitudinal momentum components, as well as all-order corrections for 
momentum configurations with large scale hierarchies are discussed below.

Let us illustrate the calculation using the red paths 
in Fig~\ref{fig:histories}. One definite path (from dashed red through solid red to the
top layer, e.g. following the rightmost lines in the figure) will contribute to the overall probability as 
\begin{eqnarray}
\label{eq:calcP}
\mathcal{P}_{\textnormal{H}}=
\me{hj}{(1)}
\otimes
P_{\textnormal{H}(1)}
&\otimes&
\overbrace{\left[\frac{\me{\mathrm{H}jj}{} }{\me{hj}{(1)}P_{\textnormal{H}(1)}+\me{hj}{(2)}P_{\textnormal{H}(2)}}\right]}^{\mc{R}(\mathrm{H}jj)}\\
\otimes
P_{\textnormal{H}}
&\otimes&\left[
\frac{\me{\gamma\gamma jj}{} }{\me{\mathrm{H}jj}{}\mc{R}(\mathrm{H}jj)P_{\textnormal{H}} + \me{\gamma\gamma j}{}\mc{R}(\gamma\gamma j)P_{\textnormal{QED}} + \me{\gamma j j}{}\mc{R}(\gamma j j)P_{\textnormal{QCD}}}
\right]~\nonumber,
\end{eqnarray}
where $P_{\textnormal{X}}$ are approximate transition kernels, for example given by
dipole splitting functions~\cite{Gustafson:1987rq,Catani:1996vz}.
In order to construct this
probability for the case of Fig~\ref{fig:histories}, 
splitting functions for all QCD and QED vertices, as well as for
Higgs-gluon, Higgs-fermion and Higgs-photon couplings have been calculated.
When summing over the two dashed red paths, the full
$\me{\mathrm{H}jj}{}$ is recovered, while summing over the dashed green and
dashed blue paths yield the full mixed QCD/QED matrix elements
$\me{\gamma\gamma j}{}$ and $\me{\gamma jj}{}$, respectively. The total sum
of the probabilities of all paths reduces to $\me{\gamma\gamma jj}{}$, as desired.
This discussion is complicated significantly by phase-space constraints, but
can be generalized to an arbitrary multiplicity and to arbitrary 
splittings~\cite{Fischer:2017yja}.

Note that it is straightforward to ``tag" a path of recombinations as QCD-, QED-
or Higgs-type by simply examining the intermediate configuration. The sum of
all probabilities of all Higgs-type paths is an excellent measure of how
Higgs-like the input state was, while the sum of all non-Higgs-type 
probabilities is an excellent measure of how background-like the input was.
Following Eq.~\ref{eq:chi}, it is thus natural to define the probability of the Higgs-hypothesis as
\begin{eqnarray}
\chi_{\textnormal{H}} \equiv \frac{\mathcal{P}(\{p_i\}|\,\textnormal{Higgs})}{\mathcal{P}(\{p_i\}|\,\neg\,\textnormal{Higgs})},
\label{eq:chih}
\end{eqnarray}
where the respective probabilities are defined as 
\begin{equation}
\mathcal{P}(\{p_i\}|\,\textnormal{Higgs}) = \frac{\sum\mathcal{P}_{\textnormal{H}}}{\sum(\mathcal{P}_{\textnormal{H}} + \mathcal{P}_{\textnormal{QCD}}+ \mathcal{P}_{\textnormal{QED}}) }
~~~ \mathrm{and} ~~~
\mathcal{P}(\{p_i\}|\,\neg\,\textnormal{Higgs}) = \frac{\sum (\mathcal{P}_{\textnormal{QCD}}+\mathcal{P}_{\textnormal{QED}})}{ \sum(\mathcal{P}_{\textnormal{H}} + \mathcal{P}_{\textnormal{QCD}}+ \mathcal{P}_{\textnormal{QED}}) }.
\label{eq:probs}
\end{equation} 
A plethora of tags defining a hypothesis can be envisioned -- once all paths 
of all intermediate states leading to the highest-multiplicity (input) state 
are known, it is straightforward to attribute a probability to each hypothesis. 
Of course, not all hypotheses are sensible from the quantum-mechanical 
perspective if interference effects are important. In this case, we expect
that if the hypothesis is tested on pseudo-data with the \maps\ method, the 
results are similar, irrespective of how the pseudo-data was generated. There
should not be strong discrimination power for such problematic hypotheses. 

Finally, a discrimination based
on matrix elements alone is likely to give an unreasonable probability
for multi-jet hadronic states, since e.g.\ large hierarchies in jet transverse momenta
will not be described by fixed-order matrix elements alone, and because the 
overall flux of initial-state partons is tied to changes in the parton
distribution functions. Thus, we
include the all-order effects of the evolution between intermediate states
into the probability of each path. For a path $p$ of intermediate states
$S_i^{(p)}, i\in[1,n^{(p)}]$ transitioning to the next higher multiplicity at scales
$t_i^{(p)}$, this can be done by correcting the 
probability of each path to
\begin{eqnarray}
\label{eq:calcWP}
\mathcal{P}_{\textnormal{A}}\rightarrow
\mathcal{P}_{\textnormal{A}} w_{p}\quad\textnormal{where}\quad
w_{p} = \prod_{i=1}^{n^{(p)}}
             \Pi(S_{i-1}^{(p)}; t_{i-1}^{(p)},t_i^{(p)})
             \, \frac{\alpha(S_i^{(p)},t_i^{(p)})}{ \alpha^{\textnormal{\tiny FIX}}(S_i^{p})}
             \, \frac{f(S_{i-1}^{(p)}; x_{i-1}^{(p)},t_{i-1}^{(p)})}
                     {f(S_{i-1}^{(p)}; x_{i-1}^{(p)},t_{i}^{(p)})} 
\end{eqnarray}
where $\Pi(S_{i-1}^{(p)}; t_{i-1}^{(p)},t_i^{(p)})$ is the no-branching
probability of state $S_{i-1}^{(p)}$ between scales $t_{i-1}^{(p)}$ and 
$t_i^{(p)}$, which is directly related to Sudakov form factors~\cite{Sudakov:1954sw,Sjostrand:1985xi}. We have also introduced the placeholder 
$\alpha^{\textnormal{\tiny FIX}}(S_i^{p})$
for the coupling constant of the branching producing state $S_i^{p}$ out of 
state $S_{i-1}^{(p)}$, and $\alpha(S_i^{(p)},t_i^{(p)})$ as a placeholder
for the same coupling evaluated taking the kinematics of state $S_i^{p}$
into account\footnote{For details on running coupling choices, see
Appendix~\ref{app:direqed}.}.
Finally, the parton luminosity appropriate for state $S_{i-1}^{(p)}$, evaluated at
longitudinal momentum fraction $x_{i-1}^{(p)}$ and factorization scale
$t_{i-1}^{(p)}$ are collected in the factors $f(S_{i-1}^{(p)}; x_{i-1}^{(p)},t_{i-1}^{(p)})$. Ratios
of these factors account for the rescaling of the initial flux due to branchings.
The weights $w_p$ are also a key component of the CKKW-L algorithm, 
which employs trial showers to generate the no-branching probabilities, and 
attaches the PDF- and $\alpha_s$ ratios as event weight to pretabulated 
fixed-order input events. 

In \maps, we also invoke trial showers to generate the
no-branching factors, i.e.\ the calculation of the weights $w_p$ is performed by 
directly using a realistic
parton shower, specifically the \dire plugin to \textsc{Pythia}.
To correctly calculate $w_{p}$ for all possible paths, we extend this parton shower to include
QED radiation (so that the shower can give a sensible all-order QED-resummed
weight for the green paths in Fig~\ref{fig:histories}) and to allow the
transitions $q\rightarrow qH, g\rightarrow g H$ and $H\rightarrow\gamma\gamma$
(in order to correctly assign the red clustering paths in 
Fig.~\ref{fig:histories}). Details on these improvements, and on the use
of matrix-element corrections in \dire, are given in Appendix~\ref{app:direqed}.

\section{Application to $\mathrm{H} \to \gamma \gamma$~+ jets}
\label{sec:results}

To assess the performance of our approach in separating signal from background,
and to showcase the scope of its potential applications, we study the signal 
process $pp \to \mathrm{H}jj$ with subsequent decay of the Higgs boson into 
photons, $\mathrm{H} \to \gamma \gamma$, at a center-of-mass energy of $\sqrt{s} = 13$ 
TeV. This process is of importance in studying the quantum numbers of the 
Higgs boson, e.g. its couplings to other Standard Model
particles~\cite{Corbett:2015mqf,Englert:2015hrx, Englert:2017aqb,Ellis:2018gqa} 
or its CP properties \cite{Plehn:2001nj,Englert:2012ct,Englert:2012xt,Bernlochner:2018opw, Englert:2019xhk}. 
Just like for the Higgs discovery channel with inclusive number of jets, 
$pp \to (\mathrm{H} \to \gamma \gamma) + X$, this channel suffers from a large 
Standard-Model continuum background. We generate signal and background events 
using MadGraph for the hard process cross section and \pythia for showering 
and hadronisation. At the generation level, we apply minimal cuts for the 
photons ($p_{T,\gamma} \geq 20$ GeV, $|\eta| < 2.5$ and 
$\Delta R_{\gamma \gamma} \geq 0.2$), and on the final state partons 
$j$ ($p_{T,j} \geq 30$ GeV, $|\eta| \leq 4.5$ and $\Delta R_{jj} \geq 0.4$).
While we do not consider detector efficiencies for the jets, we simulate the 
detector response in the reconstruction of the photons by smearing their 
energy such that the Breit-Wigner distributed invariant mass 
$m^2_{\gamma\gamma}= (p_{\gamma,1} + p_{\gamma,2})^2$ has a width of 2 GeV after
reconstruction. Under such inclusive cuts, the signal process receives 
contributions from gluon fusion, as well as from weak-boson 
fusion~\cite{DelDuca:2001eu, Klamke:2007cu}. Standard approaches to exploit 
this signal process often rely on the application of weak boson fusion 
cuts~\cite{Rainwater:1998kj,Figy:2003nv}, which render gluon fusion 
contributions sub-dominant. Instead here, we will focus on the gluon fusion 
contributions exclusively, aiming to apply \maps\ to discriminate the 
continuum di-photon background from the gluon-fusion induced Higgs 
signal\footnote{Various approaches have been proposed to reliably separate 
gluon-fusion from weak-boson fusion in this channel \cite{Rainwater:1998kj,Englert:2012ct,Andersen:2012kn}.}.

\begin{figure}
    \subfigure[][]{
        \includegraphics[width=0.41\textwidth]{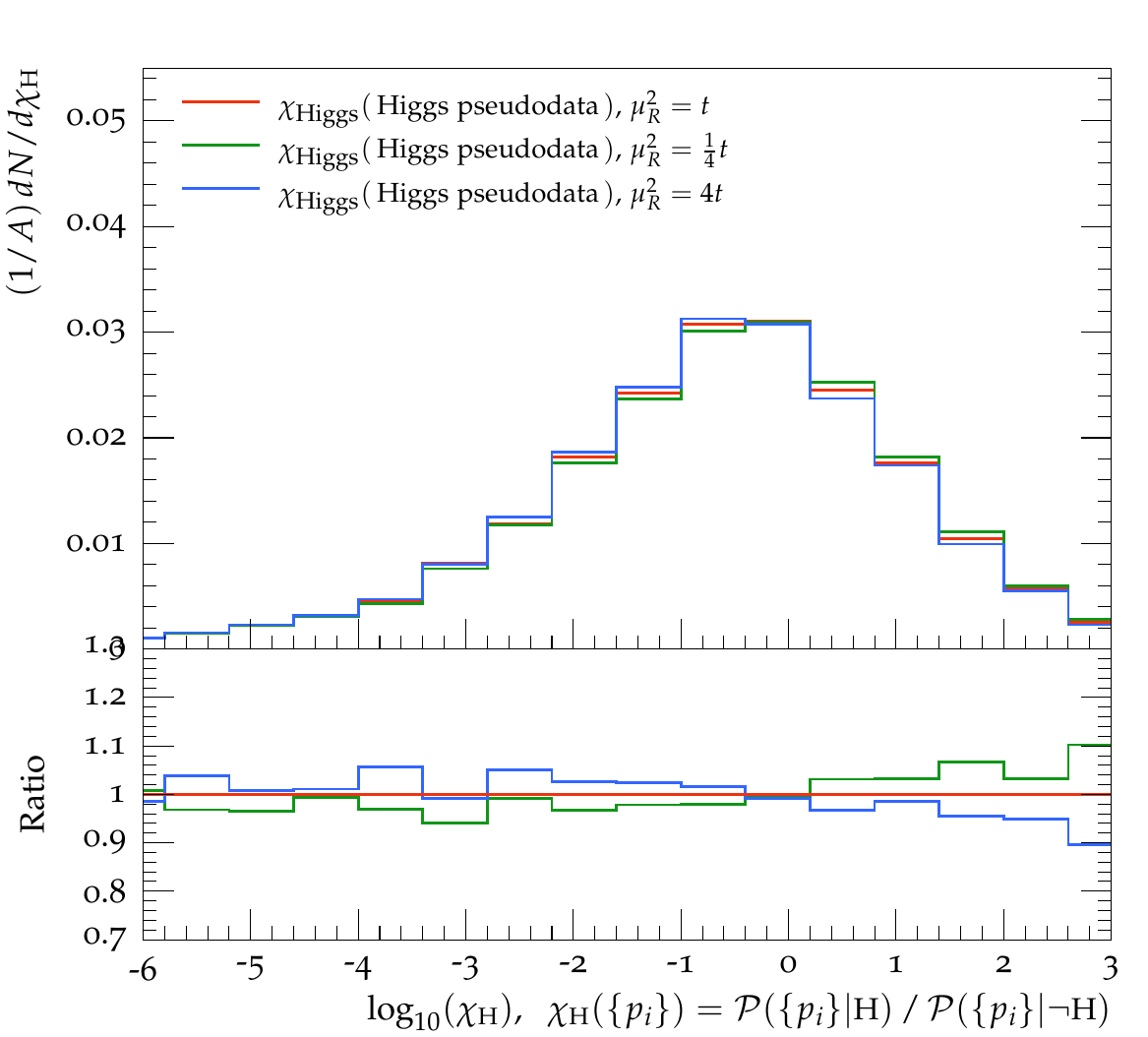}
        \label{fig:chi_signal_scale_var}
    }
    \subfigure[][]{
        \includegraphics[width=0.41\textwidth]{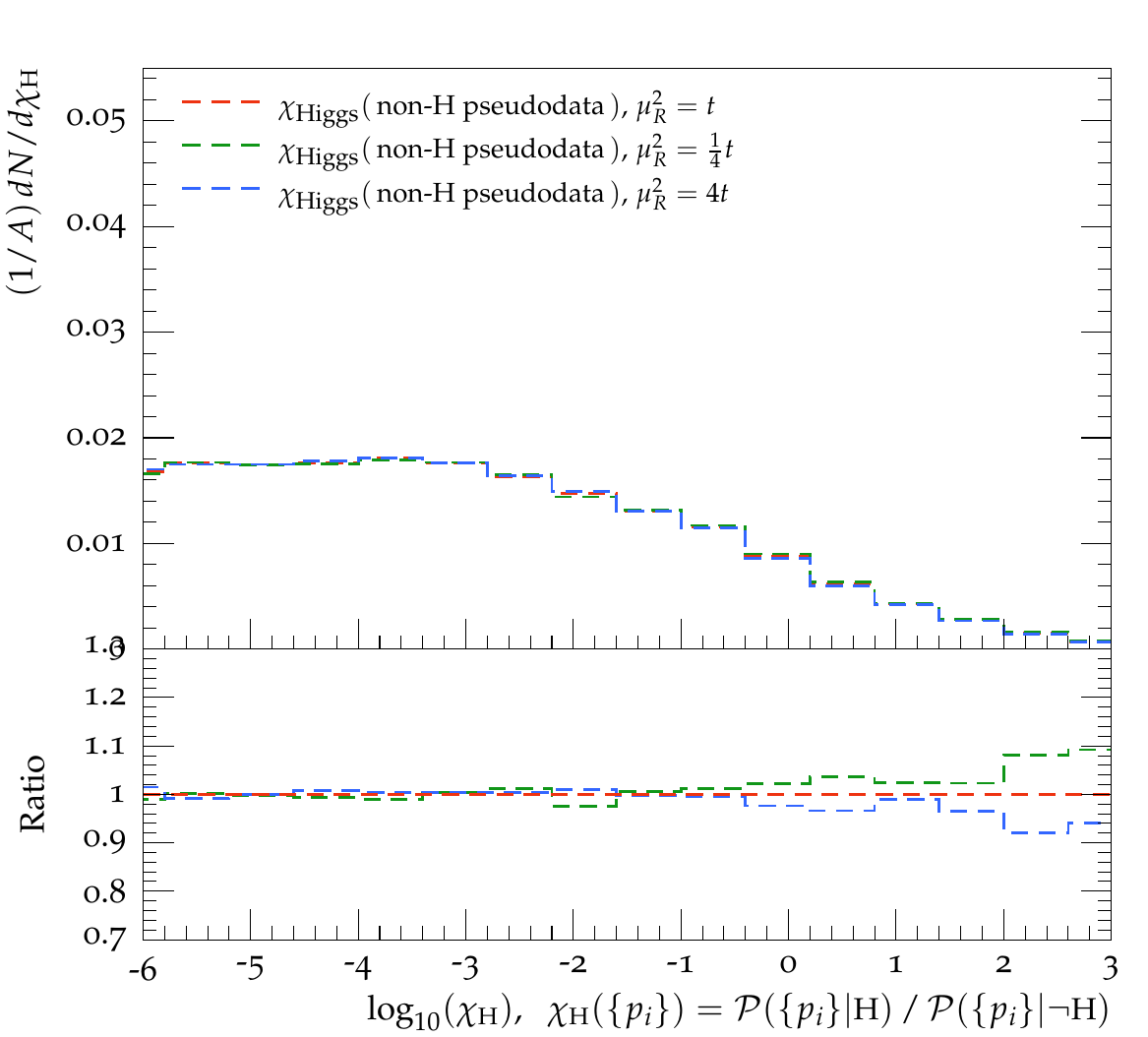}
        \label{fig:chi_bkgrnd_scale_var}
    }
\caption{
Classification of signal or background pseudodata according to Higgs
hypothesis, using different values for the argument of the QCD running coupling, both
in the evaluation of coupling factors as well as the evaluation of no-branching
probabilities.
}
\label{fig:chi_higgs_scale_var}
\end{figure}

\begin{figure}
    \subfigure[][]{
        \includegraphics[width=0.41\textwidth]{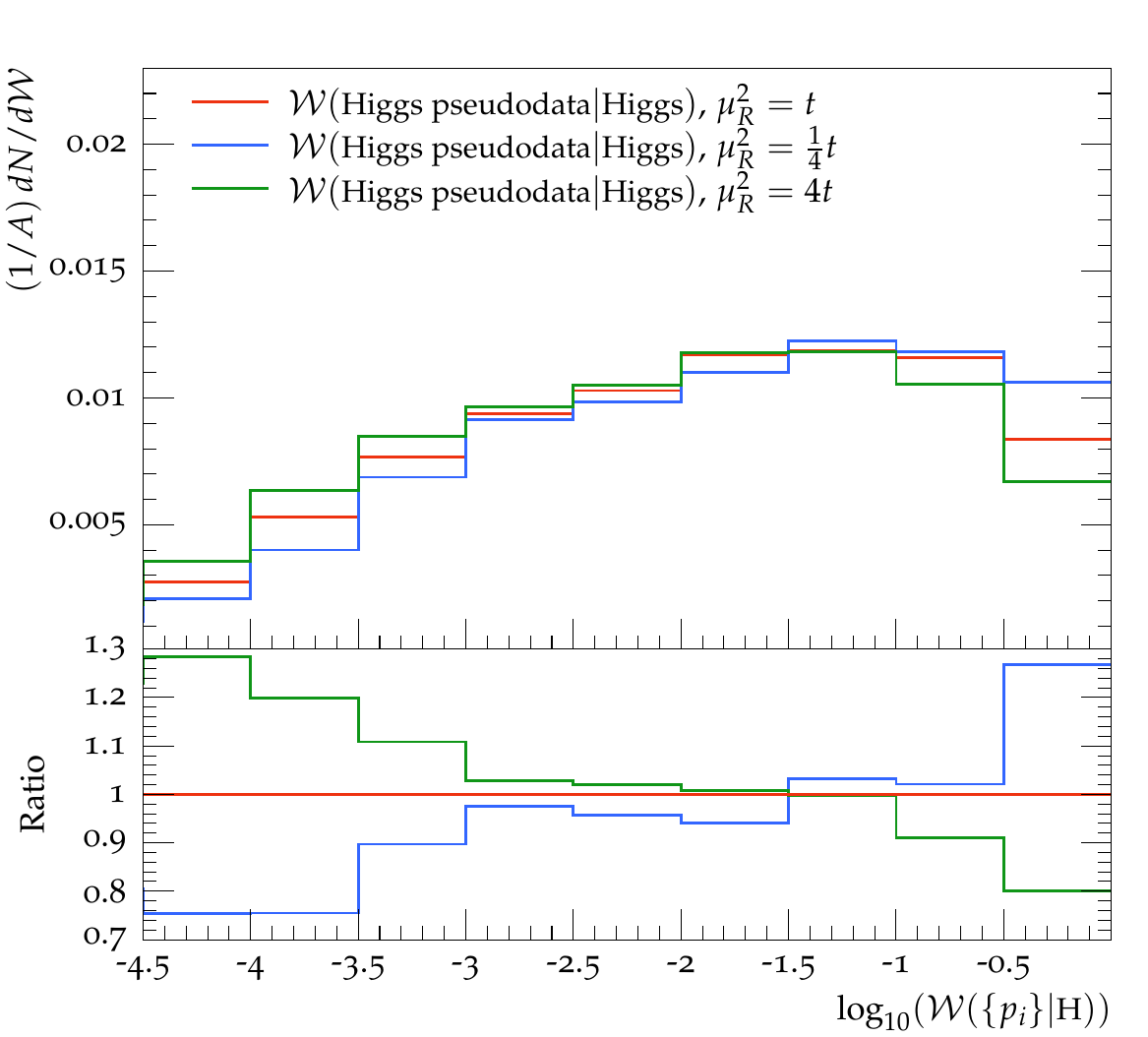}
        \label{fig:ph_signal_scale_var}
    }
    \subfigure[][]{
        \includegraphics[width=0.41\textwidth]{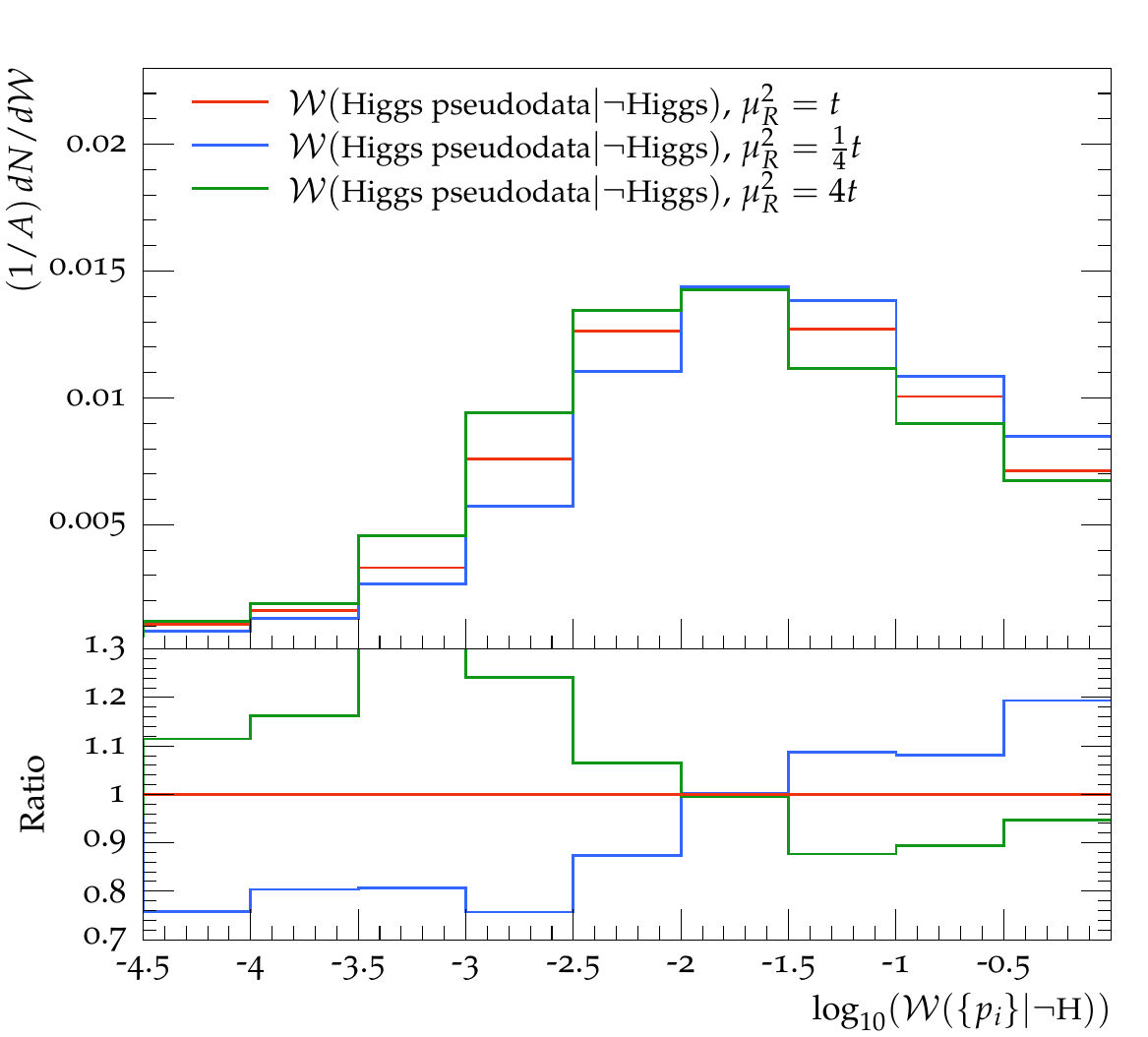}
        \label{fig:pnoth_signal_scale_var}
    }\\
    \subfigure[][]{
        \includegraphics[width=0.41\textwidth]{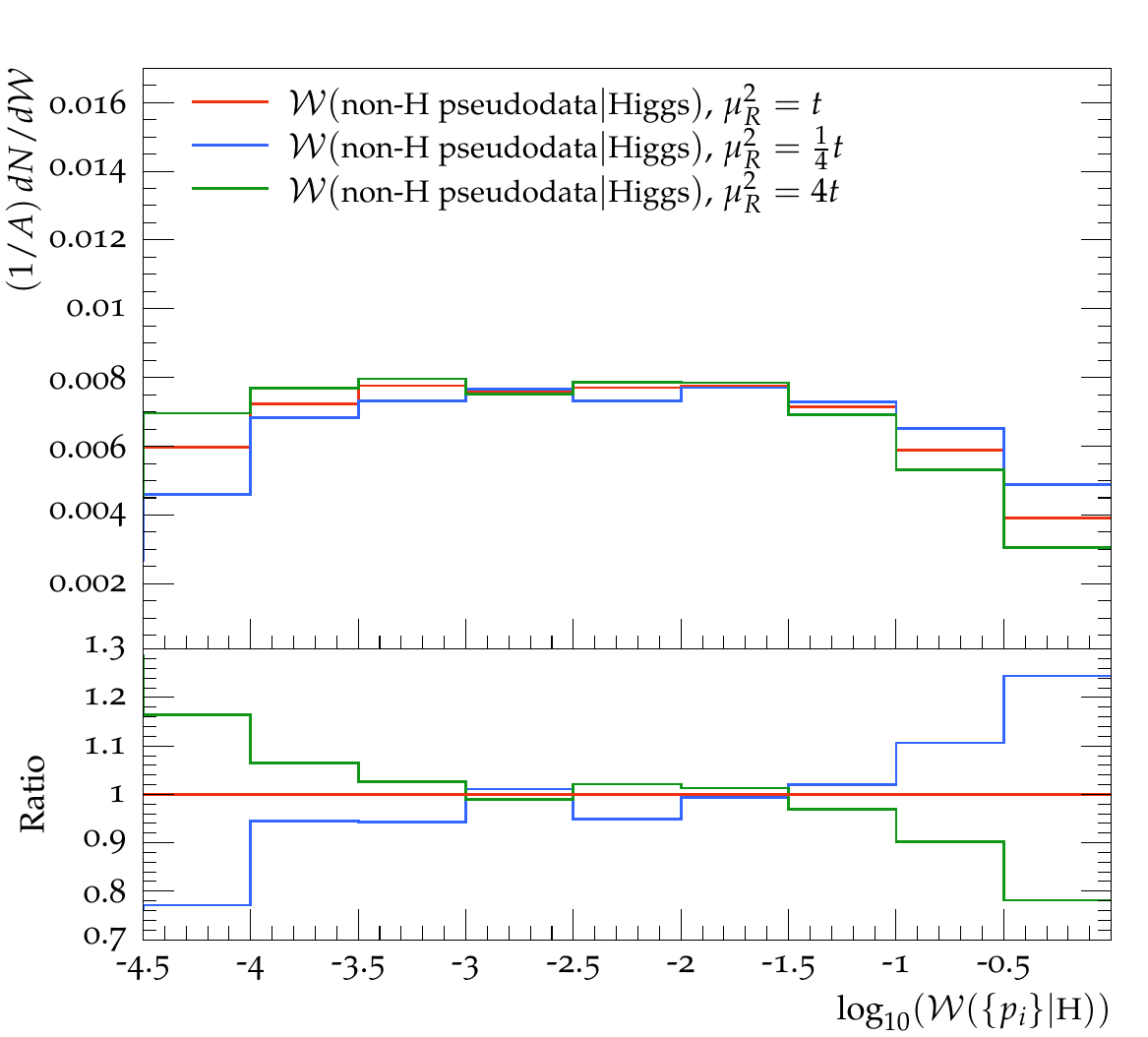}
        \label{fig:ph_bkgrnd_scale_var}
    }
    \subfigure[][]{
        \includegraphics[width=0.41\textwidth]{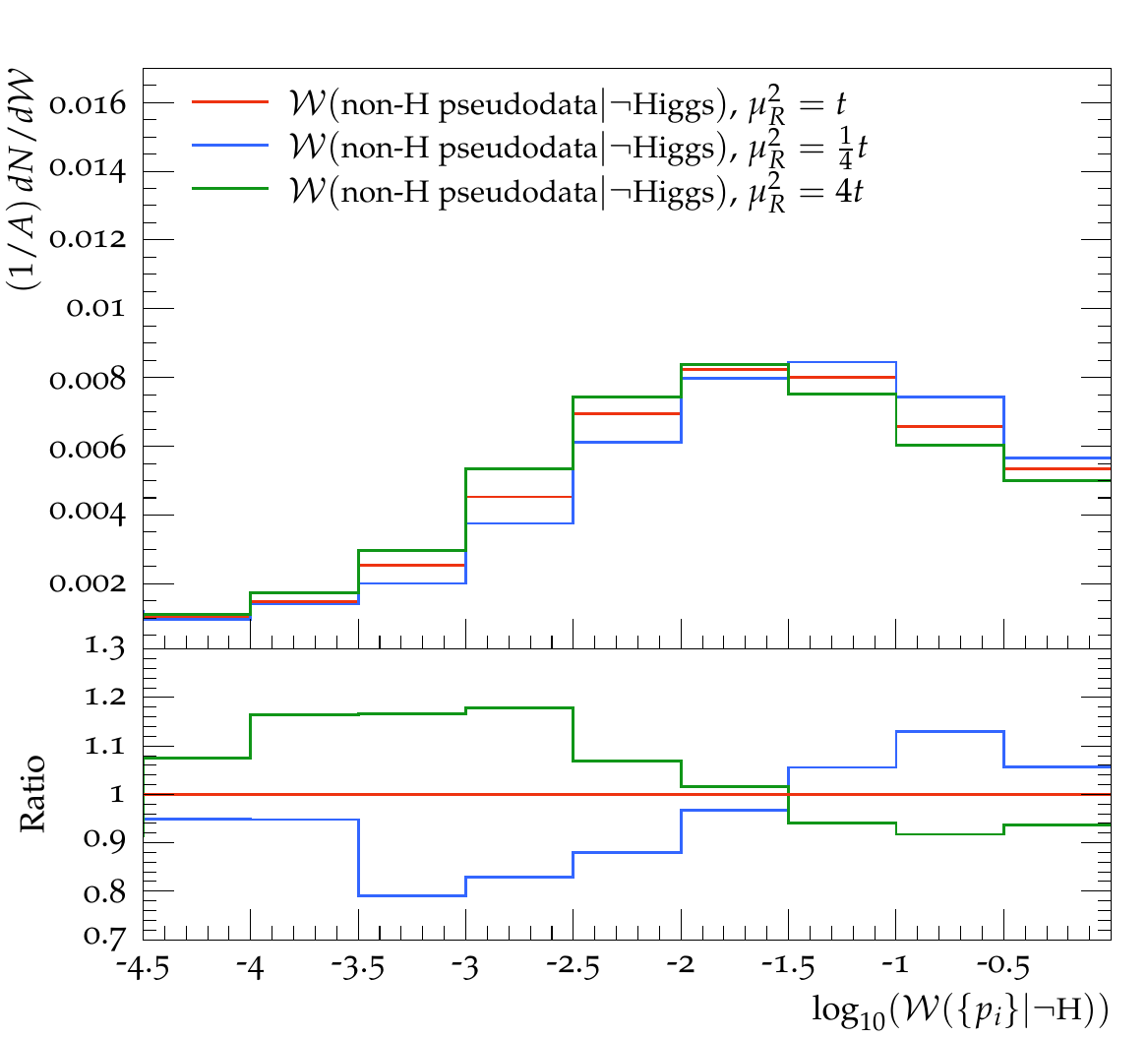}
        \label{fig:pnoth_bkgrnd_scale_var}
    }
\caption{
Non-normalized probabilities $\mathcal{W}(\{p_i\}|\,\textnormal{Hypothesis}) = 
\mathcal{P}(\{p_i\}|\,\textnormal{Hypothesis}) \cdot \sum(\mathcal{P}_{\textnormal{H}} 
+ \mathcal{P}_{\textnormal{QCD}} + \mathcal{P}_{\textnormal{QED}})$ of Higgs and 
non-Higgs pseudodata to be tagged as Higgs or non-Higgs 
configuration, using different values for the argument of the QCD running coupling, both
in the evaluation of coupling factors as well as the evaluation of no-branching
probabilities.}
\label{fig:p_higgs_scale_var}
\end{figure}

In Fig.~\ref{fig:chi_higgs_scale_var} we show $\log_{10}(\chi_{\textnormal{H}})$,
as calculated according to Eqs.~\ref{eq:chi} and ~\ref{eq:calcP}, for 
Higgs-signal pseudo-data (left) and non-Higgs background samples (right). It 
is apparent that the observable $\chi$ can discriminate between signal and 
background events. Signal events have on average large $\chi_{\textnormal{H}}$, 
i.e. they result in a relatively large value for $\mathcal{P}(\{p_i\}|S)$ in 
comparison to $\mathcal{P}(\{p_i\}|B)$, and vice versa for background events.
Since the \maps\ method is based on calculating well-defined perturbative 
factors, it goes beyond many existing classification methods by also providing 
an estimate of theoretical uncertainties of the hypothesis-testing 
variable $\chi_{\textnormal{H}}$. We find that the theoretical uncertainty, 
estimated by varying the renormalisation scale between $t/2 \leq \mu_R \leq 2 t$ (where 
$t$ are the \dire parton-shower evolution variables~\cite{Hoche:2015sya}, as 
necessary to evaluate running $\alpha_s$ effects at the nodal splittings in the 
history tree, and to perform $\mu_R$-variations of the no-branching factors) are
very small for $\chi$ in our example. This is somewhat remarkable, as signal and
background enter to lowest order at $\mathcal{O}(\alpha^2_s)$ for the hard 
process. As shown in Fig.~\ref{fig:p_higgs_scale_var}, 
$\mathcal{P}(\{p_i\}|S)$ and $\mathcal{P}(\{p_i\}|B)$ separately (and multiplied
by the total probability to ensure that no artificial numerator-denominator
cancellations occur) show a large
sensitivity on scale variations, which cancels when taking the ratio to 
calculate $\chi_{\textnormal{H}}$.
This can also be understood in terms of a cancellation for the performance of 
the classifier. In the calculation of both the signal and the background hypotheses,
partons are interpreted as emitted from the initial state partons, thus forming 
the final states with two (or more) jets. As this underlying dynamics is 
governed by QCD, this is very similar for signal and background, so that this 
part of the event does not contain much discriminative information. Furthermore,
changing the argument of $\alpha_s$ will affects signal and background in a 
similar way.

\begin{figure}
        \includegraphics[width=0.5\textwidth]{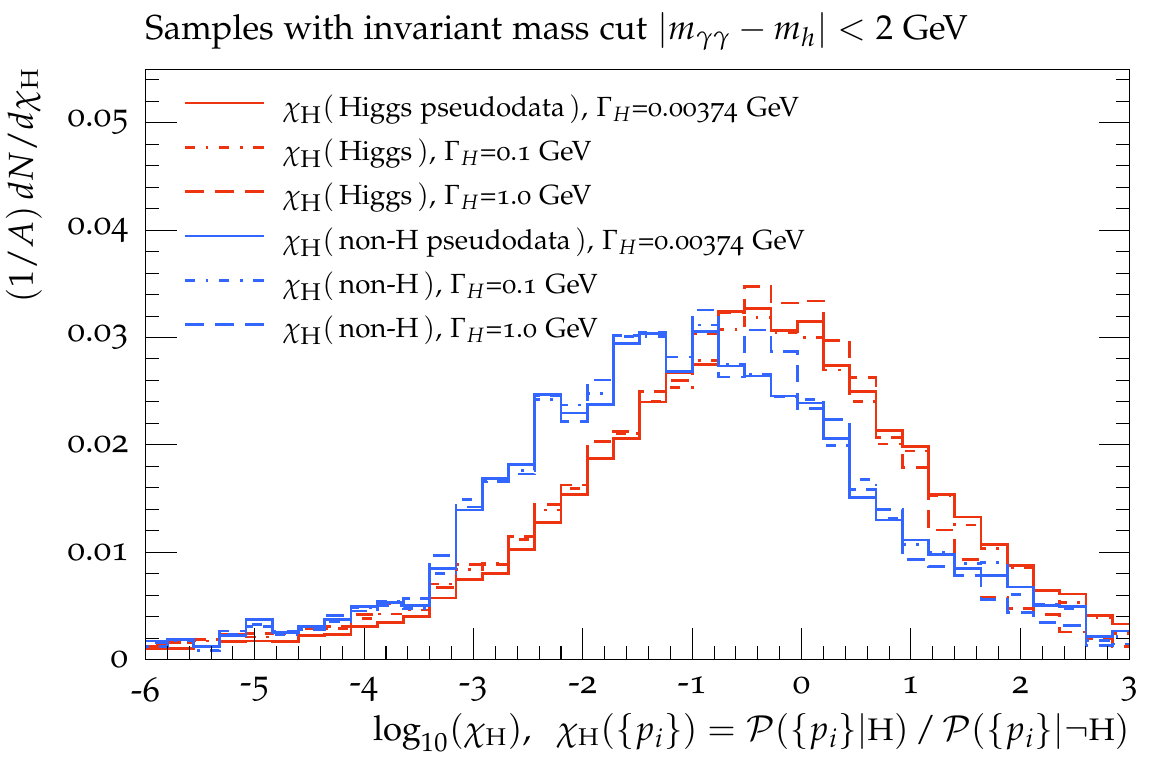}
\caption{\label{fig:chi_higgs_gamma_var}
Classification of signal or background pseudodata according to Higgs
hypothesis, using different values of $\Gamma_{H}$. Only configurations with 
diphoton invariant masses in a small window are shown, to further demonstrate
the discrimination power w.r.t. a simple mass cut.
}
\end{figure}

\begin{figure}
    \label{fig:ps}
    \subfigure[][]{
        \includegraphics[width=0.45\textwidth]{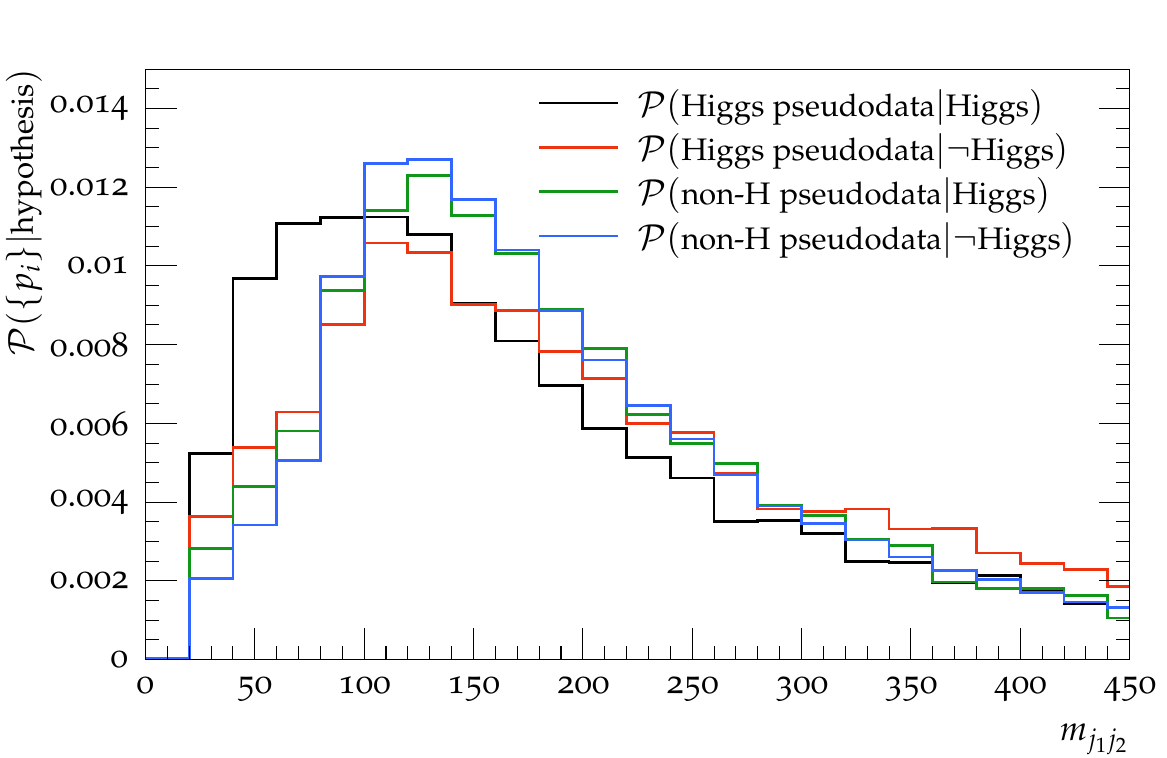}
        \label{fig:p_mjj}
    }
    \subfigure[][]{
        \includegraphics[width=0.45\textwidth]{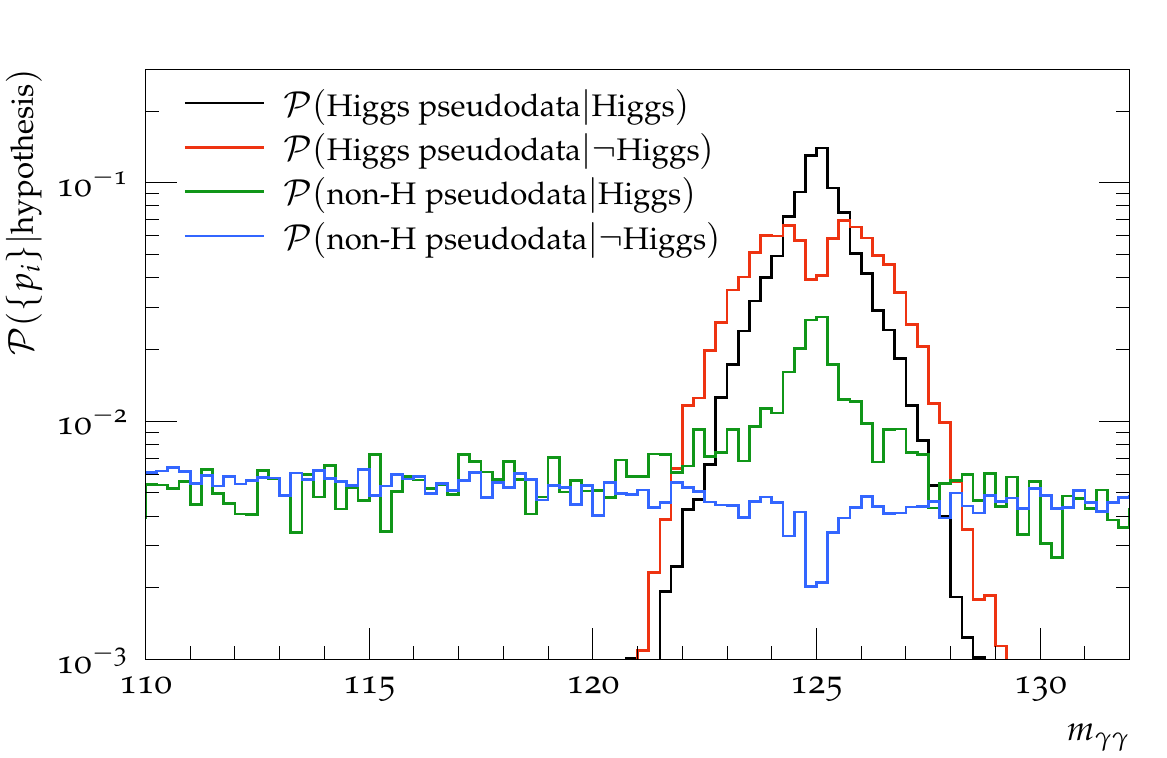}
        \label{fig:p_maa}
    }
\caption{
Probabilities to identify signal or background pseudodata according to ``Higgs-signal" and ``non-Higgs signal" hypothesis, as function
of the dijet invariant mass and the diphoton invariant mass.
}
\label{fig:probsvsmass}
\end{figure}

This raises the question whether all information used in discriminating signal 
from background is in fact contained in the electroweak part of the event, and 
could e.g. be captured by analyzing the invariant mass 
distribution $m_{\gamma \gamma}$. We can investigate the effect of a 
mass-window cut within experimental uncertainties by selecting signal and 
background events that satisfy 
$|m_{\gamma \gamma} - 125~\mathrm{GeV}| < 2~\mathrm{GeV}$, in line with the way
we smeared the energy of the photons. Fig.~\ref{fig:chi_higgs_gamma_var} shows 
when applying a mass cut, the normalised distributions of $\chi_{\textnormal{H}}$ overlap much 
more for signal and background samples, indicating that the very good 
separating observed in Fig.~\ref{fig:chi_higgs_scale_var} rests largely on the 
fact that the photons in the signal arise due to the decay of a narrow 
resonance. Still, the signal samples result on average in a large value 
for $\chi_{\textnormal{H}}$ compared to the background samples and 
thus $S/B$ can be improved with a cut on $\chi_{\textnormal{H}}$. 

In order to construct the history tree for the \maps\ method, it was necessary
to introduce ``Higgs splitting kernels" (cf.~App.~\ref{app:dire}) to define
the probability of the $\textnormal{H}\rightarrow \gamma\gamma$ decay. In 
principle, it would be permissible to use the physical Higgs-boson width when 
calculating these splitting kernels. However, it is reasonable to expect that
this might lead to an artificially strong discrimination power. 
Fig.~\ref{fig:chi_higgs_gamma_var} shows that this is not the case, by
varying the Higgs-boson width in the splitting kernel in a very large range.

The \maps\ method effectively takes all possible observables into account 
to discriminate between two hypotheses. To investigate further how this
relates to cutting on $m_{\gamma \gamma}$,  Fig.~\ref{fig:probsvsmass} shows 
the probabilities $\mathcal{P}$ directly, binned in the differential 
distributions $m_{\gamma \gamma}$ and $m_{jj}$. This highlights that \maps\
might also be useful to find optimal cuts in a cut-and-count analysis, since
\maps\ can quantify how much differential observables can discriminate between 
different hypotheses. As shown in Fig.~\ref{fig:probsvsmass}, $m_{jj}$ is very
similar for signal and backgrounds, while $m_{\gamma \gamma}$ is very 
discriminative. The sensitivity of any observable in classifying events can
be studied in this way.

\begin{figure}
    \subfigure[][]{
        \includegraphics[width=0.31\textwidth]{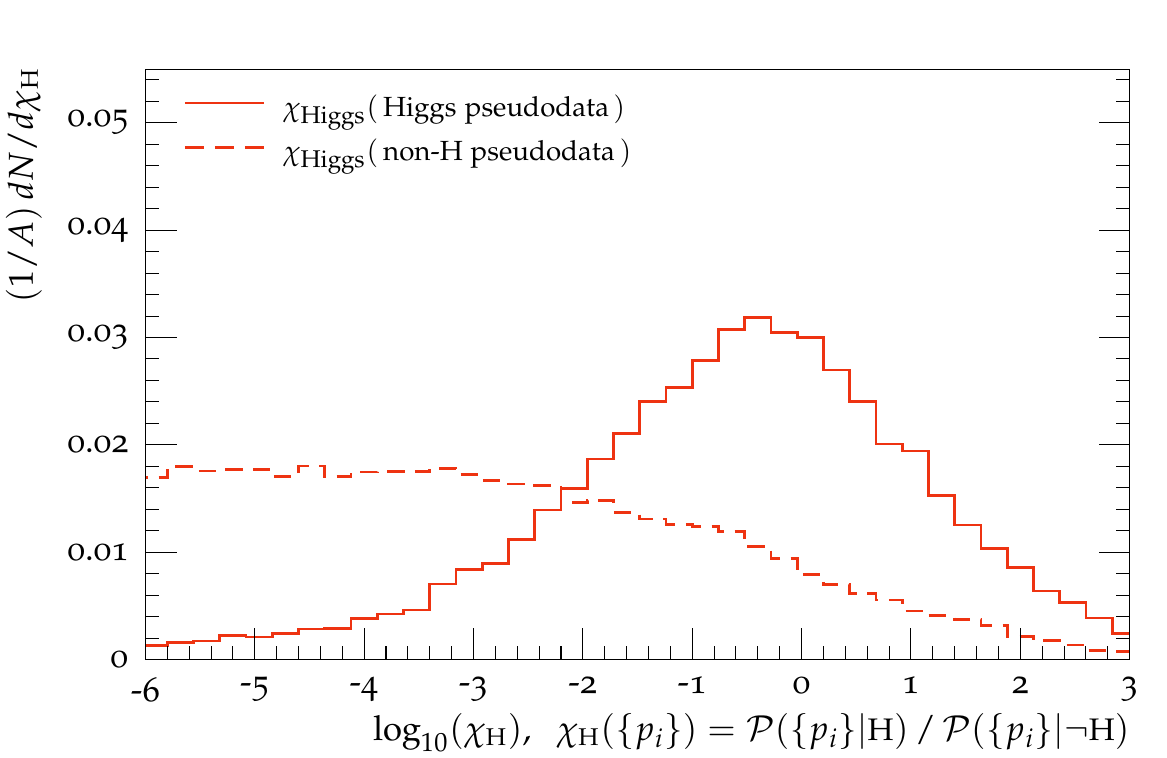}
        \label{fig:chi_higgs}
    }
    \subfigure[][]{
        \includegraphics[width=0.31\textwidth]{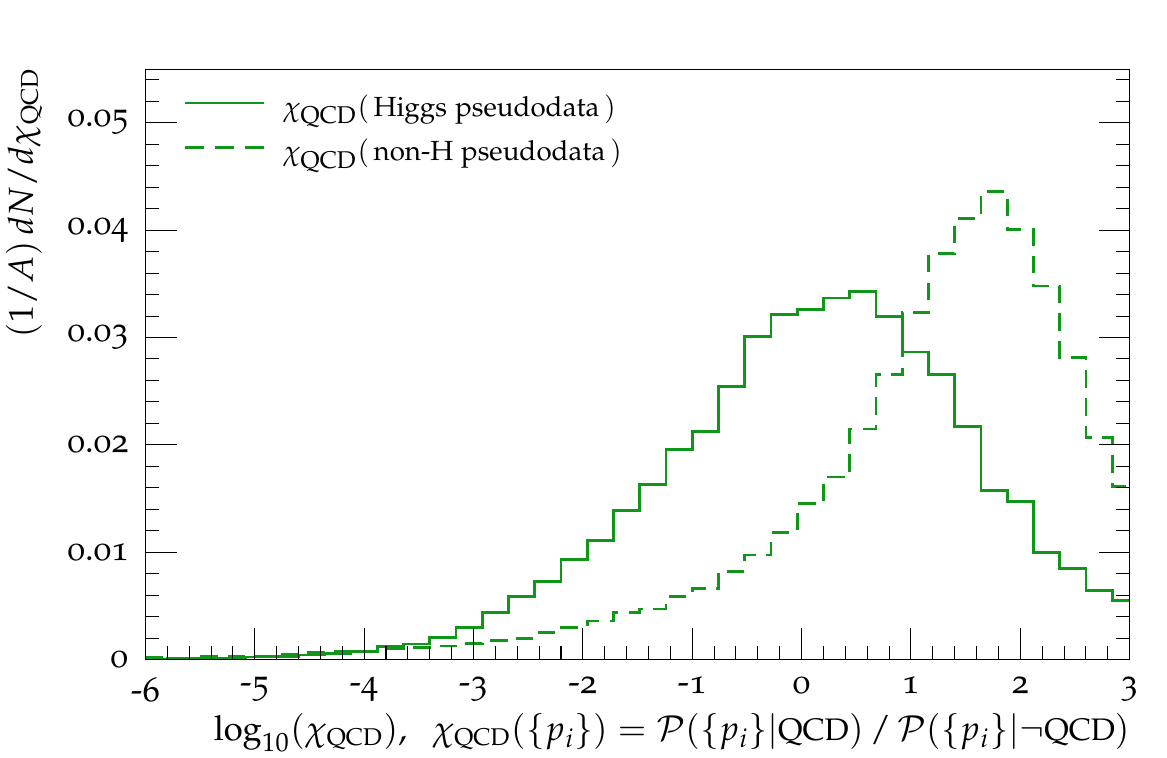}
        \label{fig:chi_qcd}
    }
    \subfigure[][]{
        \includegraphics[width=0.31\textwidth]{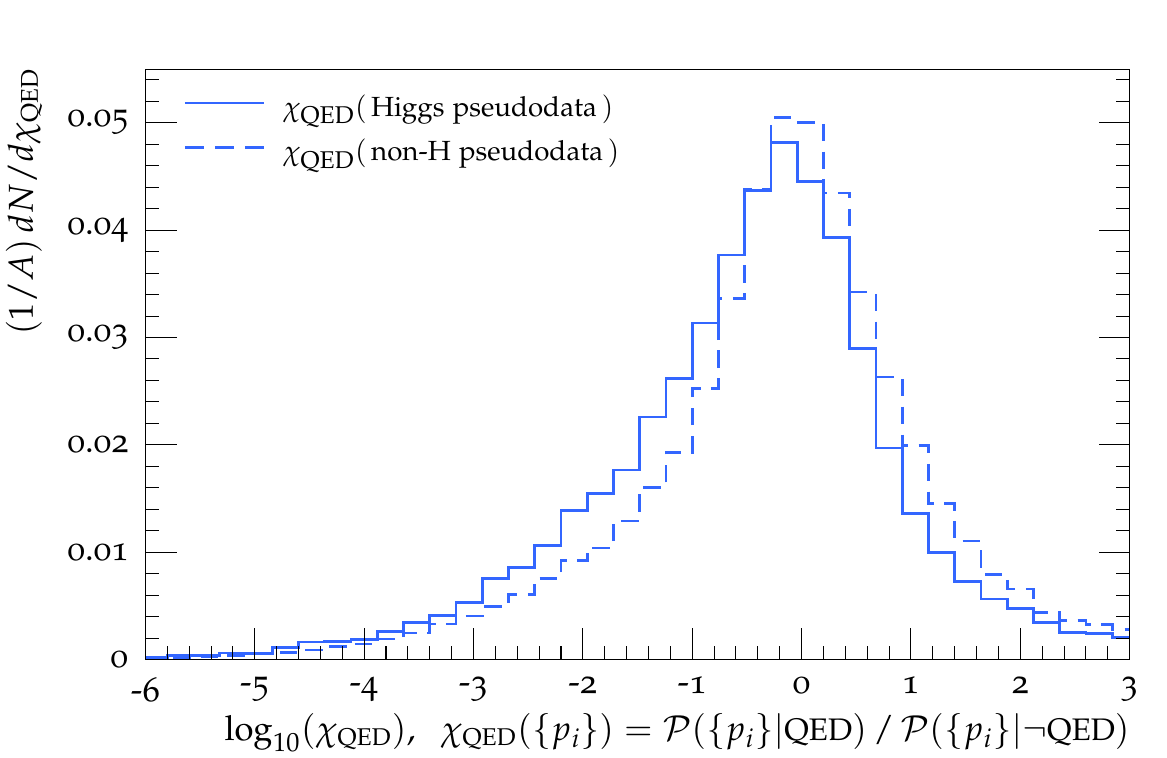}
        \label{fig:chi_qed}
    }
\caption{
Classification of signal or background pseudodata according to different
signal hypotheses. Panel (a): Higgs hypothesis $\chi_{\textnormal{H}}$ ($\Gamma_H=0.1$ GeV), Panel (b): QCD hypothesis $\chi_{\textnormal{QCD}}$, 
Panel (c): QED hypothesis $\chi_{\textnormal{QED}}$.
}
\label{fig:chis}
\end{figure}

Classification with respect to Higgs or no-Higgs hypotheses is not the only 
application for \maps\ in our example. One can imagine to construct different 
classification observables to test different hypotheses. For example, we could
define $\chi_{\textnormal{QED}}$ and $\chi_{\textnormal{QCD}}$ in analogy to 
Eqs.~(\ref{eq:chih}) and (\ref{eq:probs}), i.e.
\begin{eqnarray}
\chi_{\textnormal{QED}} \equiv \frac{\mathcal{P}(\{p_i\}|\,\textnormal{QED})}{\mathcal{P}(\{p_i\}|\,\neg\,\textnormal{QED})}
\qquad\textnormal{and}  \qquad
\chi_{\textnormal{QCD}} \equiv \frac{\mathcal{P}(\{p_i\}|\,\textnormal{QCD})}{\mathcal{P}(\{p_i\}|\,\neg\,\textnormal{QCD})},
\end{eqnarray}
with the probabilities 
\begin{eqnarray}
\mathcal{P}(\{p_i\}|\,\textnormal{QED}) &= \frac{\sum\mathcal{P}_{\textnormal{QED}}}{\sum(\mathcal{P}_{\textnormal{H}} + \mathcal{P}_{\textnormal{QCD}}+ \mathcal{P}_{\textnormal{QED}}) }, 
~~~~\mathcal{P}(\{p_i\}|\,\neg\,\textnormal{QED}) = \frac{\sum (\mathcal{P}_{\textnormal{QCD}}+\mathcal{P}_{\textnormal{H}})}{ \sum(\mathcal{P}_{\textnormal{H}} + \mathcal{P}_{\textnormal{QCD}}+ \mathcal{P}_{\textnormal{QED}}) }\\ 
\mathcal{P}(\{p_i\}|\,\textnormal{QCD}) &= \frac{\sum\mathcal{P}_{\textnormal{QCD}}}{\sum(\mathcal{P}_{\textnormal{H}} + \mathcal{P}_{\textnormal{QCD}}+ \mathcal{P}_{\textnormal{QED}}) },
~~~~\mathcal{P}(\{p_i\}|\,\neg\,\textnormal{QCD}) = \frac{\sum (\mathcal{P}_{\textnormal{QED}}+\mathcal{P}_{\textnormal{H}})}{ \sum(\mathcal{P}_{\textnormal{H}} + \mathcal{P}_{\textnormal{QCD}}+ \mathcal{P}_{\textnormal{QED}}) }.
\end{eqnarray}
In Fig.~\ref{fig:chis}, we show how the Higgs-signal and non-Higgs background 
samples fare regarding these three classification variables 
$\chi_\textnormal{H}$, $\chi_\mathrm{QED}$ and $\chi_\mathrm{QCD}$. The best 
discrimination between signal and background is observed in $\chi_\textnormal{H}$. This is 
not surprising, as $\chi_\textnormal{H}$ tests explicitly if there is a Higgs boson in the 
sample or not. $\chi_\mathrm{QCD}$ and $\chi_\mathrm{QED}$ perform as expected,
yielding an on average larger value of $\chi$ for the background sample, and 
smaller values for the events that do contain a Higgs boson. While $\chi_\mathrm{QCD}$ 
retains some discriminative power between the Higgs and no-Higgs samples, the 
least discriminate variable is $\chi_\mathrm{QED}$. Hence, with respect to the 
green path in Fig.~\ref{fig:histories}, the signal and background samples 
provide very little separable kinematic features. The $\mathrm{QED}$ hypothesis
provides a very similar classifier, irrespective of the event sample, indicating
that no ``classical" path in the history tree is preferred, and that
thus, interferences are relevant. It is comforting that in this case, the \maps\
method does indeed, as desired, not produce an artificial discrimination
power by referring to classical paths.
In conclusion, by applying \maps~to known signal and background samples it is 
possible to optimise the discriminating observable, and to obtain an improved 
understanding of the kinematic features that allow a discrimination between 
signal and backgrounds.

\section{Conclusions}
\label{sec:conclusion}
The classification of events into signal and background is the basis for all 
searches and measurements at collider experiments. By building on the Event 
Deconstruction method \cite{Soper:2011cr, Soper:2014rya}, CKKW-L 
merging \cite{Lonnblad:2001iq} and the iterated matrix-element correction 
approach of \cite{Fischer:2017yja}, we have developed and implemented a novel
way to classify realistic (i.e. fully showered and hadronised) final states according to 
different theory hypotheses. This method has been implemented in a standalone 
package, called \maps, and will be made publicly available.

In principle this method is applicable to any final state and any theoretical 
hypotheses. However, there is a practical limitation due to the sharply 
increasing time it takes to evaluate complex final states with many 
(colored) particles. While invisible particles have not been implemented 
yet, approaches how to take them into account in the hypothesis testing 
exist~\cite{FerreiradeLima:2017iwx} and will be included in a future release of \maps.

We have applied \maps~to the gluon-fusion induced production of H$jj$ with 
subsequent decay H$ \to \gamma \gamma$. This process receives large backgrounds
where the photons can either be produced in the hard interaction of the 
process $pp \to \gamma \gamma jj$ or by being radiated off the final state or 
initial state quarks of the process $pp \to jj$. Detector effects were rudimentarily 
taken into account by smearing the photon momenta. \maps~can directly calculate 
the probability of how likely an event was produced through a transition of 
interest. We have shown that \maps~can confidently separate between signal 
and background samples with respect to the Higgs or no-Higgs hypothesis. While 
the method takes into account all possible kinematic observables simultaneously
to classify the event according to the hypotheses of consideration, it is 
also possible to study how much individual observables, or combinations of 
observables, contribute to the overall classification. Thus, \maps~can be used 
to optimise cuts for cut-and-count based analyses very efficiently. The 
flexible and first-principle calculation-based approach enables us to obtain an 
improved understanding of the kinematic features that allow us to discriminate 
between signal and backgrounds for very large classes of processes at any 
high-energy collider experiment.

\section{Acknowledgments}
We thank Valentin Hirschi for collaboration during an early stage of this 
project, by sharing a private code to generate all color 
connections, and for longstanding help with using
MadGraph to generate the C++ matrix element code employed for matrix
element corrections. MS is grateful to Dave Soper for a longstanding collaboration
on the Shower/Event Deconstruction approach. MS thanks the University of 
Tuebingen and the Humboldt Society for support and hospitality during the 
finalisation of parts of this work. SP would like to thank Walter Giele for 
collaboration on dipole showers for QED splittings.

\appendix 

\section{QCD, QED and Higgs splittings in the \dire dipole shower}\label{app:direqed}
\label{app:dire}

Realistic classifications of final states containing jets and photons 
according to an hypothesis require the construction of all possible branching
histories that could have produced the final states. Thus, all possible ways
of splitting or recombining the particles in the final state have to be 
considered. For the problem at hand, this requires a simultaneous description 
of QCD- and QED branchings, both at fixed- and all-order perturbative 
accuracy. If an hypothesis does not only depend on the final-state
particles alone, but rather infers reconstructed intermediate state such
as Higgs bosons, it is also necessary to incorporate the relevant 
intermediate branchings.

The description of QCD splittings used in this publication is implemented in 
the \dire plugin to \pythia, and consists of a 
partial-fractioned dipole parton shower including mass effects, as documented 
in detail in~\cite{Hoche:2015sya}. For all QCD splittings, we evaluate
the running coupling at the evolution scale assigned to the splitting, 
i.e.~$\alpha(S_i^{(p)},t_i^{(p)}) = \alpha_{s}(t_i^{(p)})$.

We implement QED  emissions as an extension of the partial fractioned
dipole shower of \dire~\cite{Hoche:2015sya}, using the same evolution and energy sharing
variables as well as kinematical splitting functions (and mass corrections)
as discussed in~\cite{Hoche:2015sya}. The crucial difference to the treatment of QCD is
that we allow all pairs of electric charges to form dipoles that coherently
emit photons, similar to the ideas presented in~\cite{Kniehl:1992ra} and
more recently discussed in~\cite{Schonherr:2017qcj} and~\cite{Kleiss:2017iir}. 
At variance to the latter, we split the soft-photon radiation pattern into
two pieces each assigned to one dipole splitting kernel. The color factors 
in the QCD splitting functions in~\cite{Hoche:2015sya} are further replaced by 
the electric (dipole) charge correlators, which can readily be negative. This 
inconvenience is addressed by using the weighted parton shower~\cite{Hoeche:2009xc,Lonnblad:2012hz} 
algorithm implemented in \dire. The assignment of recoilers for the 
$\gamma\rightarrow f\bar f$ splitting takes guidance from the simultaneous 
emission of a soft quark pair in QCD (see e.g.~\cite{Catani:1999ss}) which 
can be thought of being emitted from a parent color 
dipole~\cite{Dulat:2018vuy}. The latter calculation is of course not directly
applicable to QED. Nevertheless, in the absense of other concrete ideas,
we allow all electrically charged particles to
act as spectator for the $\gamma\rightarrow f\bar f$ splitting.
For all QED branchings, we do not employ a running
coupling and instead fix $\alpha$ to the Thompson value (cf.~\cite{Kniehl:1992ra}.), i.e.\
$\alpha(S_i^{(p)},t_i^{(p)}) = \alpha_{em}(0) = 0.00729735$. More details
on the formalism of QED showers will be presented elsewhere~\cite{Giele:2019xxx}.

Since our QED splitting kernels can readily become negative, we expect that
the event weight fluctuation due to the weighting algorithm can become a 
significant problem. This is however largely circumvented by including QCD and QED
matrix element corrections up to $pp\rightarrow \gamma\gamma j j j$ in
the formalism of ~\cite{Fischer:2017yja} into the parton shower: Since the 
matrix-element corrections guarantee the correct radiation pattern irrespective
of the splitting kernels, it is legitimate to enforce positive splitting kernels
for splittings yielding states for which matrix-element corrections are 
available, thus not producing large weight fluctuations.

To allow testing the hypothesis of an intermediate Higgs boson, we further
include the emission rate $g\rightarrow g$H and the decay rate 
H$\rightarrow \gamma\gamma$ directly into the parton shower evolution.
The emission rate $q\rightarrow q$H is omitted, since its contribution is only
present for heavy quarks and is, due to the quark masses, further suppressed by
phase space.
The evolution variable and phase space mapping for the emission rate 
$g\rightarrow g$H is identical to that of (massive) QCD or QED splittings, and 
the splitting function is a simple uniform weight 
$\Gamma_{\textnormal{H}\rightarrow gg}(m_\mathrm{H})$. This allows to assign a probability to
the production vertex of the Higgs boson, and is sufficient as long as the 
emission rate is effectively absent in the shower evolution.
All gluons that can be reached by tracing leading-$N_C$ color connections are 
possible spectators for this splitting.
The coupling value $\alpha(S_i^{(p)},t_i^{(p)})$ for the $g\rightarrow g$H
emission is fixed to $\alpha(S_i^{(p)},t_i^{(p)})=\Gamma_{\mathrm{H}\rightarrow gg}(m_\mathrm{H})$

The virtuality of the photon pair serves as evolution variable for the 
H$\rightarrow \gamma\gamma$ decay. In this case, the splitting kernel is
defined by
\begin{eqnarray*}
P_{\mathrm{H}\rightarrow \gamma\gamma} = \frac{1}{\mathcal{S}}
\Gamma_{\mathrm{H}\rightarrow \gamma\gamma} 
\frac{8\pi p_\mathrm{H}^2}{ (p_\mathrm{H}^2 - m_\mathrm{H}^2)^2 + p_\mathrm{H}^2\Gamma^2_{tot~\mathrm{H}} }
\end{eqnarray*}
where $\mathcal{S}$ is the number of possible recoilers for this splitting.
In line with the reasoning for the $\gamma\rightarrow f\bar f$ splitting above,
we allow all gluons as spectators for this splitting. Again, it worth
noting that we do employ matrix-element corrections for shower
splittings that produce $pp\rightarrow \gamma\gamma j j j$  or less complicated
states, such that for the purposes of this publication, the concrete 
prescription of the $P_{\mathrm{H}\rightarrow \gamma\gamma}$ is of minor importance. 
The coupling value $\alpha(S_i^{(p)},t_i^{(p)})$ for the 
H$\rightarrow \gamma\gamma$ decay is fixed to 
$\alpha(S_i^{(p)},t_i^{(p)})=\Gamma_{\mathrm{H}\rightarrow \gamma\gamma}(m_\mathrm{H})$

\bibliography{ref}{}

\begin{thebibliography}{10}

\bibitem{James:2000et}
F.~James, Y.~Perrin, and L.~Lyons, editors,
\newblock {\em {Workshop on confidence limits, CERN, Geneva, Switzerland, 17-18
  Jan 2000: Proceedings}}, 2000.

\bibitem{Kondo:1988yd}
K.~Kondo,
\newblock J. Phys. Soc. Jap. {\bf 57}, 4126 (1988).

\bibitem{Abazov:2004cs}
D0, V.~M. Abazov {\em et~al.},
\newblock Nature {\bf 429}, 638 (2004), hep-ex/0406031.

\bibitem{Abulencia:2005pe}
CDF, A.~Abulencia {\em et~al.},
\newblock Phys. Rev. {\bf D73}, 092002 (2006), hep-ex/0512009.

\bibitem{Cranmer:2006zs}
K.~Cranmer and T.~Plehn,
\newblock Eur. Phys. J. {\bf C51}, 415 (2007), hep-ph/0605268.

\bibitem{Gao:2010qx}
Y.~Gao {\em et~al.},
\newblock Phys. Rev. {\bf D81}, 075022 (2010), 1001.3396.

\bibitem{Andersen:2012kn}
J.~R. Andersen, C.~Englert, and M.~Spannowsky,
\newblock Phys. Rev. {\bf D87}, 015019 (2013), 1211.3011.

\bibitem{Martini:2015fsa}
T.~Martini and P.~Uwer,
\newblock JHEP {\bf 09}, 083 (2015), 1506.08798.

\bibitem{Gritsan:2016hjl}
A.~V. Gritsan, R.~R{\"o}ntsch, M.~Schulze, and M.~Xiao,
\newblock Phys. Rev. {\bf D94}, 055023 (2016), 1606.03107.

\bibitem{Soper:2011cr}
D.~E. Soper and M.~Spannowsky,
\newblock Phys. Rev. {\bf D84}, 074002 (2011), 1102.3480.

\bibitem{Soper:2012pb}
D.~E. Soper and M.~Spannowsky,
\newblock Phys. Rev. {\bf D87}, 054012 (2013), 1211.3140.

\bibitem{FerreiradeLima:2016gcz}
D.~Ferreira~de Lima, P.~Petrov, D.~Soper, and M.~Spannowsky,
\newblock Phys. Rev. {\bf D95}, 034001 (2017), 1607.06031.

\bibitem{Soper:2014rya}
D.~E. Soper and M.~Spannowsky,
\newblock Phys. Rev. {\bf D89}, 094005 (2014), 1402.1189.

\bibitem{Englert:2015dlp}
C.~Englert, O.~Mattelaer, and M.~Spannowsky,
\newblock Phys. Lett. {\bf B756}, 103 (2016), 1512.03429.

\bibitem{FerreiradeLima:2017iwx}
D.~E. Ferreira~de Lima, O.~Mattelaer, and M.~Spannowsky,
\newblock Phys. Lett. {\bf B787}, 100 (2018), 1712.03266.

\bibitem{Buckley:2011ms}
A.~Buckley {\em et~al.},
\newblock Phys. Rept. {\bf 504}, 145 (2011), 1101.2599.

\bibitem{Mangano:2001xp}
M.~L. Mangano, M.~Moretti, and R.~Pittau,
\newblock Nucl. Phys. {\bf B632}, 343 (2002), hep-ph/0108069.

\bibitem{Catani:2001cc}
S.~Catani, F.~Krauss, R.~Kuhn, and B.~R. Webber,
\newblock JHEP {\bf 11}, 063 (2001), hep-ph/0109231.

\bibitem{Lonnblad:2001iq}
L.~L{\"o}nnblad,
\newblock JHEP {\bf 05}, 046 (2002), hep-ph/0112284.

\bibitem{Giele:2011cb}
W.~T. Giele, D.~A. Kosower, and P.~Z. Skands,
\newblock Phys. Rev. {\bf D84}, 054003 (2011), 1102.2126.

\bibitem{Fischer:2017yja}
N.~Fischer and S.~Prestel,
\newblock Eur. Phys. J. {\bf C77}, 601 (2017), 1706.06218.

\bibitem{Hoche:2015sya}
S.~H{\"o}che and S.~Prestel,
\newblock Eur. Phys. J. {\bf C75}, 461 (2015), 1506.05057.

\bibitem{Dobbs:2001ck}
M.~Dobbs and J.~B. Hansen,
\newblock Comput. Phys. Commun. {\bf 134}, 41 (2001).

\bibitem{Alwall:2014hca}
J.~Alwall {\em et~al.},
\newblock JHEP {\bf 07}, 079 (2014), 1405.0301.

\bibitem{Sjostrand:2014zea}
T.~Sj{\"o}strand {\em et~al.},
\newblock Comput. Phys. Commun. {\bf 191}, 159 (2015), 1410.3012.

\bibitem{Cacciari:2008gp}
M.~Cacciari, G.~P. Salam, and G.~Soyez,
\newblock JHEP {\bf 04}, 063 (2008), 0802.1189.

\bibitem{Cacciari:2011ma}
M.~Cacciari, G.~P. Salam, and G.~Soyez,
\newblock Eur. Phys. J. {\bf C72}, 1896 (2012), 1111.6097.

\bibitem{Alwall:2006yp}
J.~Alwall {\em et~al.},
\newblock Comput. Phys. Commun. {\bf 176}, 300 (2007), hep-ph/0609017.

\bibitem{Gustafson:1987rq}
G.~Gustafson and U.~Pettersson,
\newblock Nucl. Phys. {\bf B306}, 746 (1988).

\bibitem{Catani:1996vz}
S.~Catani and M.~H. Seymour,
\newblock Nucl. Phys. {\bf B485}, 291 (1997), hep-ph/9605323,
\newblock [Erratum: Nucl. Phys.B510,503(1998)].

\bibitem{Sudakov:1954sw}
V.~V. Sudakov,
\newblock Sov. Phys. JETP {\bf 3}, 65 (1956),
\newblock [Zh. Eksp. Teor. Fiz.30,87(1956)].

\bibitem{Sjostrand:1985xi}
T.~Sj{\"o}strand,
\newblock Phys. Lett. {\bf 157B}, 321 (1985).

\bibitem{Corbett:2015mqf}
T.~Corbett {\em et~al.},
\newblock (2015), 1511.08188.

\bibitem{Englert:2015hrx}
C.~Englert, R.~Kogler, H.~Schulz, and M.~Spannowsky,
\newblock Eur. Phys. J. {\bf C76}, 393 (2016), 1511.05170.

\bibitem{Englert:2017aqb}
C.~Englert, R.~Kogler, H.~Schulz, and M.~Spannowsky,
\newblock Eur. Phys. J. {\bf C77}, 789 (2017), 1708.06355.

\bibitem{Ellis:2018gqa}
J.~Ellis, C.~W. Murphy, V.~Sanz, and T.~You,
\newblock JHEP {\bf 06}, 146 (2018), 1803.03252.

\bibitem{Plehn:2001nj}
T.~Plehn, D.~L. Rainwater, and D.~Zeppenfeld,
\newblock Phys. Rev. Lett. {\bf 88}, 051801 (2002), hep-ph/0105325.

\bibitem{Englert:2012ct}
C.~Englert, M.~Spannowsky, and M.~Takeuchi,
\newblock JHEP {\bf 06}, 108 (2012), 1203.5788.

\bibitem{Englert:2012xt}
C.~Englert, D.~Goncalves-Netto, K.~Mawatari, and T.~Plehn,
\newblock JHEP {\bf 01}, 148 (2013), 1212.0843.

\bibitem{Bernlochner:2018opw}
F.~U. Bernlochner {\em et~al.},
\newblock (2018), 1808.06577.

\bibitem{Englert:2019xhk}
C.~Englert, P.~Galler, A.~Pilkington, and M.~Spannowsky,
\newblock (2019), 1901.05982.

\bibitem{DelDuca:2001eu}
V.~Del~Duca, W.~Kilgore, C.~Oleari, C.~Schmidt, and D.~Zeppenfeld,
\newblock Phys. Rev. Lett. {\bf 87}, 122001 (2001), hep-ph/0105129.

\bibitem{Klamke:2007cu}
G.~Klamke and D.~Zeppenfeld,
\newblock JHEP {\bf 04}, 052 (2007), hep-ph/0703202.

\bibitem{Rainwater:1998kj}
D.~L. Rainwater, D.~Zeppenfeld, and K.~Hagiwara,
\newblock Phys. Rev. {\bf D59}, 014037 (1998), hep-ph/9808468.

\bibitem{Figy:2003nv}
T.~Figy, C.~Oleari, and D.~Zeppenfeld,
\newblock Phys. Rev. {\bf D68}, 073005 (2003), hep-ph/0306109.

\bibitem{Kniehl:1992ra}
B.~A. Kniehl and L.~L{\"o}nnblad,
\newblock {Renormalization scales in electroweak physics: and Photon radiation
  in the dipole model and in the Ariadne program},
\newblock in {\em {Workshop on Photon Radiation from Quarks Annecy, France,
  December 2-3, 1991}}, pp. 109--112, 1992.

\bibitem{Schonherr:2017qcj}
M.~Sch{\"o}nherr,
\newblock Eur. Phys. J. {\bf C78}, 119 (2018), 1712.07975.

\bibitem{Kleiss:2017iir}
R.~Kleiss and R.~Verheyen,
\newblock JHEP {\bf 11}, 182 (2017), 1709.04485.

\bibitem{Hoeche:2009xc}
S.~Hoeche, S.~Schumann, and F.~Siegert,
\newblock Phys. Rev. {\bf D81}, 034026 (2010), 0912.3501.

\bibitem{Lonnblad:2012hz}
L.~L{\"o}nnblad,
\newblock Eur. Phys. J. {\bf C73}, 2350 (2013), 1211.7204.

\bibitem{Catani:1999ss}
S.~Catani and M.~Grazzini,
\newblock Nucl. Phys. {\bf B570}, 287 (2000), hep-ph/9908523.

\bibitem{Dulat:2018vuy}
F.~Dulat, S.~H{\"o}che, and S.~Prestel,
\newblock Phys. Rev. {\bf D98}, 074013 (2018), 1805.03757.

\bibitem{Giele:2019xxx}
W.~Giele and S.~Prestel,
\newblock publication in preparation .

\end{thebibliography}
\bibliographystyle{h-physrev}

\end{document}